\documentclass[preprint,showkeys,amsmath,amssymb]{revtex4}
\usepackage{graphicx}
\usepackage{dcolumn}
\usepackage{bm}

\begin{document}


\title{Dark energy and Dark matter interaction in light of the second law of thermodynamics}
\author{V\'ictor H. C\'ardenas}
\email{vitor.cardenas@uv.cl}
\author{Daniela Grand\'on}
\email{dgrandons@gmail.com}
\affiliation{Instituto de F\'{\i}sica y Astronom\'ia, Universidad de
Valpara\'iso, Gran Breta\~na 1111, Valpara\'iso, Chile.}
\author{Samuel Lepe}
 \email{samuel.lepe@pucv.cl}
\affiliation{Instituto de F\'{\i}sica, Pontificia Universidad Cat\'olica de Valpara\'iso, Av. Brasil 2950, Valpara\'iso, Chile}

\date{\today}

\begin{abstract}
In the context of thermodynamics we discuss the way inevitable emerge an interaction between dark components, and in this way, provide a mechanism to understand the limits of the LCDM model and the class of interaction models between dark components. Using observational data we have tested two models of explicit interaction between dark components and reconstructed the evolution of temperatures for both components. We found that observations suggest the interaction exist with energy flowing from dark energy to dark matter. The best fit also suggest a phantom equation of state parameter for dark energy. We discuss the results having in mind the constraints imposed by thermodynamics.
\end{abstract}

\keywords{Dark energy; Dark Matter; Interaction}
\maketitle

\section{Introduction}

One of the main problems in cosmology today is to identify or
characterize the cause of the accelerated expansion found in
\cite{ref:Riess98,ref:Perlmuter99}, usually called dark energy (DE).
Although there have been, from time to time, evidence for DE
evolution, or in its primordial form, evidence for a variable
cosmological constant, the case have taken new impulse after the
results using the BAO BOSS DR11 \cite{Delubac:2014aqe}. In that
work, by interpreting the results in a simple wCDM model, strongly
suggest a $2.5 \sigma$ departure from LCDM at $z=2.34$. 
Soon after this work appears, several papers focused on finding what 
model better describe the results from the BOSS analysis. 
In \cite{Sahni:2014ooa} the authors found evidence for DE evolution
using the $H(z)$ measurement implied by \cite{Delubac:2014aqe}. They
propose a model where the cosmological constant was screened in the
past. Also in \cite{Abdalla:2014cla} the authors demonstrate that it 
is possible to explain the BOSS result in the context of the interacting 
dark scenario (dark matter (DM) interacting with DE), excluding null 
interaction at $2\sigma$. Also in \cite{Ding:2015vpa} the authors found 
similar results to \cite{Sahni:2014ooa}. Regarding the dark interaction 
scenario, using data from Planck, type Ia supernova, and redshift space 
distortions, in \cite{Salvatelli:2014zta} the authors found statistical 
evidence for a dark interacting model (the so called Interacting vacuum 
model, named iVCDM , \cite{wands}) starting at $z\simeq 0.9$.

One of the main problems with dark interaction models is the
arbitrariness in the coupling, usually named $Q$. In particular, in
the iVCDM the coupling is written as $Q=-q_v H \rho_v$, where $H$ is the
Hubble function, $q_v(z)$ is the arbitrary constant usually parameterized 
in bins to be reconstructed by observations, and $\rho_v$ is the vacuum energy 
density. In some sense, the same pathological feature $\Lambda$ has in 
the $\Lambda$CDM model, here emerges through $Q$ in the interacting dark 
matter/energy model. So it would be interesting to find ways to constraint 
$Q$ not only through observations but also from physical principles.

In this paper, we want to explore to what extent, the interaction
term $Q$ can be determined and constrained by appealing to thermodynamic
considerations. In what follows we use natural units, $c=8\pi G=k_B=1$. 
The paper is organized as follows: in the next section we discuss the thermodynamics in the context of an interacting model. In section III we discuss the method of effective equation of state and obtain the expressions for the temperature as a function of redshift. Then we describe the models of interaction we study and the data we have used to constraint the models in section V. We end with the discussion of our results.

\section{Thermodynamics and interaction}

\subsection{A single fluid}

Let us start with general considerations about thermodynamics in an expanding 
Universe using a single fluid. Assuming a flat universe and a homogeneous 
component we get the Hubble equation and the conservation equation
\begin{equation}\label{conseq1}
    3H^2 = \rho, \hspace{1cm} \dot{\rho} + 3H(\rho + p)=0, 
\end{equation}
where $H=\dot{a}/a$ is the Hubble parameter and $a(t)$ the scale factor. From 
the second law of thermodynamics applied to a comoving volume element of unit 
coordinate volume (and physical volume $V=a^3$) we can write
\begin{equation}\label{tds1}
TdS = d\left[(\rho + p) V\right] - Vdp.
\end{equation}
Here $\rho$ is the energy density, $p$ is the pressure, $T$ the temperature 
of the system and $S$ the total entropy per comoving volume. Using the fact 
that $S$ is an state function
we get
\begin{equation}
\frac{\rho + p}{T} = \frac{dp}{dT},
\end{equation}
then Eq.(\ref{tds1}) can be written as
\begin{equation}
dS = \frac{d\left[(\rho + p) V\right]}{T} - V (\rho + p) \frac{dT}{T^2}=d\left[ \frac{(\rho + p) V}{T}\right],
\end{equation}
which means that entropy can be written as
\begin{equation}\label{sconst}
S = \text{const.} + \frac{(\rho + p) a^3}{T}.
\end{equation}
On the other hand, from Einstein's equations we get the energy conservation equation
(\ref{conseq1}) that can be written as
\begin{equation}
d\left[(\rho + p) V\right] = Vdp,
\end{equation}
which means that the entropy (\ref{sconst}) is constant during the expansion,
\begin{equation}\label{svalue}
S = \frac{(\rho + p) a^3}{T}=\text{const.}
\end{equation}
This results emphasize the {\it adiabatic expansion} universe picture. Then Einstein's
equations implies adiabatic expansion assuming $T \neq 0$. Also, Eq.(\ref{svalue}) 
tell us that the special combination of physical variables keep a constant value once 
the evolution of $p(a)$, $\rho(a)$ and $T(a)$ are introduced. Clearly, once we know 
the equation of state (EoS), we can infer the temperature evolution of the system.

Another way to see this, is by 
using that our single component satisfies number and energy conservation. Following
\citep{Maartens:1996vi}
it is possible to find the relation
\begin{equation}\label{tpunto}
\dot{T} = -3H T \frac{\partial p}{\partial \rho}.
\end{equation}
This equation also enable us to find how the temperature evolves if an EoS is given. For example, for $p = \omega \rho$ we get
\begin{equation}
T(a) \simeq a^{-3\omega},
\end{equation}
that gives the expected answer for radiation ($\omega_r = 1/3$). The same result can be obtained from Eq.(7). 
%
In general, assuming $\omega = \omega(a)$ we can write
\begin{equation}\label{tda}
T(a) = T(a_0) \exp \left( -3 \int_{a_0}^{a} da \frac{\omega(a)}{a}\right).
\end{equation}

\subsection{Two fluids}

Let us assume now two fluids -- DM and DE for example -- and assume that both 
components conserved separately, i.e. they satisfy
\begin{eqnarray} \label{conseq}
\dot{\rho}_{x}+3H\left( 1+\omega _{x}\right) \rho _{x} &=& 0, \\
\dot{\rho}_{m}+3H\left( 1+\omega _{m}\right) \rho _{m} &=& 0,
\end{eqnarray}
so no interaction is present. For each component is possible to write the same 
equations derived before. In particular, we can write the evolution of temperature 
for these two components as
\begin{equation}
T_{m} \simeq \frac{1}{a^2}, \hspace{1cm} T_{x} \simeq  \exp \left( -3 \int da \frac{\omega_{x}}{a}\right), 
\end{equation}
where we have used the previous result for dust. It is clear that this result suggest 
that today $T_{m} \ll T_{x}$ because $\omega_{x} < 0$, as was first notice in 
\cite{Pavon:2007gt}. It is also clear that in this case the expansion is adiabatic, 
because each one of the conservation equations above can be written as $dU + pdV = 0$ 
where $U = \Omega_v \rho a^3 $ and $V = \Omega_v a^3$ where $\Omega_v$ is a volume 
constant factor.

The current observational evidence -- so far -- support the $\Lambda$CDM model, where
$\Lambda$ is a component that can be considered as a fluid with EoS parameter 
$\omega=-1$, plus a cold (non relativistic velocities) DM component, both of which 
conserved separately, i.e., they do not interact. In this context we can ask, what 
temperature can be associated to this $\Lambda$ fluid? From our previous considerations,
assuming $p=\omega \rho$ with $\omega = \text{const.}$ we can write
\begin{equation}
dS=\left( 1+\omega \right) d\left( \frac{\rho V}{T}\right),
\label{(9)}
\end{equation}
from which we can obtain
\begin{equation}
\left( 1+\omega \right) \rho V = \text{const.} \times T.  \label{(10)}
\end{equation}
This relation suggest that for a fluid with $\omega=-1$ we get $T=0$, which according
to the third law of thermodynamics implies $S=0$. This result -- independent of the 
awkward features of existence of a pure cosmological constant component -- tells us 
that such a component is at least inconsistent with thermodynamic too. 

Moreover, for the case $\omega <-1$? i.e., the phantom case, following the same 
argument we get $T<0$, and if we use the Euler relation 
\begin{equation}
TS=\left( 1+\omega \right) \rho V\Longrightarrow TS<0\rightarrow
S>0, \label{(11)}
\end{equation}%
which is the well known phantom problem. In this regard in \cite{phantomp} a solution is proposed in the context of irreversible thermodynamics.


Now, let us turn on the interaction between the dark components. The conservation 
equations can then be written as
\begin{eqnarray}\label{rox}
\dot{\rho}_{x}+3H\left( 1+\omega _{x}\right) \rho _{x} &=&-Q, \\ \label{rom}
\dot{\rho}_{m}+3H\left( 1+\omega _{m}\right) \rho _{m} &=& Q.
\end{eqnarray}
According to (\ref{tds1}) these equations tell us that the entropy for {\it each 
component} is not constant. By re-written each one as the first law we get for the 
interaction function $Q$
\begin{equation}\label{QdT}
Q = -\frac{T_x}{V}\frac{dS_x}{dt} = \frac{T_m}{V}\frac{dS_m}{dt}.
\end{equation}
%
%
No interaction $Q=0$ means both $S_m$ and $S_x$ being constants, i.e. we have an 
adiabatic evolution, in contrast to an interacting model, where a $Q \neq 0$ leads to
\begin{equation}\label{tds}
T_{m}dS_{m}+T_{x}dS_{x}=0,  
\end{equation}
then we get non-adiabaticity. By using (\ref{QdT}) we can also see this in another 
form by written
\begin{equation}
d\left( S_{m}+S_{x}\right) =-\left( \frac{T_{m}}{T_{x}}-1\right) dS_{m}\neq
0\Longrightarrow S_{m}+S_{x}\neq \text{const.} 
\end{equation}%
According to (\ref{QdT}) also we get that
\begin{equation}
Q>0\Longrightarrow dS_{x}<0\text{ \ \ }\text{and}\text{ \ \ }dS_{m}>0,  
\end{equation}
implying the DE entropy decreases as the DM entropy increases. From (\ref{tds}) it is 
direct to write
\begin{equation}
\frac{d}{dt}\left( S_{x}+S_{m}\right) =-\left( \frac{T_{m}}{T_{x}}-1\right)
\frac{dS_{m}}{dt}>0\Longrightarrow T_{m}<T_{x}, 
\end{equation}%
because $dS_{m}/dt>0$. Then at any instant DM is cooler than the DE. We can also write 
this last relation using (\ref{QdT}) to get
\begin{equation}
\frac{d}{dt}\left( S_{x}+S_{m}\right) =\left( \frac{1}{T_{m}}-\frac{1}{T_x}\right)VQ > 0 \iff Q>0 \text{    and    } T_{m}<T_{x}, 
\end{equation}

Having found these results, let us discuss a specific model. Let us study an interacting model close to the 
$\Lambda$CDM one. According to (\ref{rox}) and (\ref{rom})
\begin{eqnarray}
\dot{\rho}_{x}+3H(1+\omega _{x})\rho _{x} &=&-Q=\left( \frac{1}{V}\right)
T_{x}\frac{dS_{x}}{dt},  \label{(21)} \\
\dot{\rho}_{m}+3H(1+\omega _{m})\rho _{m} &=&Q=\left(
\frac{1}{V}\right) T_{m}\frac{dS_{m}}{dt},  \label{(22)}
\end{eqnarray}
and for the purpose in hands, let us assume $\omega _{x}=-1\left( \Lambda \right) $, 
and  $\omega _{m}=0\left( CDM\right) $, then we get
\begin{eqnarray}
\dot{\rho}_{x} &=&-Q\Longrightarrow \rho _{x}=-\int dtQ,  \label{(23)} \\
\dot{\rho}_{m}+3H\rho _{m} &=&Q,  \label{(24)}
\end{eqnarray}%
which reduces to $\Lambda$CDM for $Q=0$ zero interaction. This model is actually the 
i$\Lambda$CDM model of \cite{wands}. It was this model that was tested in \cite{Salvatelli:2014zta} against observation finding positive evidence for interaction. Given there are evidence for an evolving $\rho_x$ that means -- given the well known results in \cite{saulo} -- we have also evidence for an interacting model.  

For a component $\rho _{x}=\Lambda $ ($\omega _{x}=-1$) it is clear that $Q=0$ and then 
$T_{m}dS_{m}/dt=0$, and beacuse $T_{m}=const.$, then $S_{m}=const.$ And also as we saw 
before, because $T_{x}\left( \omega _{x}=-1\right) =0$ and $S_{x}\left( \omega_{x}=-1\right) =0.$

In conclusion, $\Lambda CDM$ is well supported by observations and also the assumption 
of adiabatic evolution ( $S=const.$) is consistent with the philosophy of the standard model 
that ensures adiabatic evolution after inflation ends. Then -- if there is evidence for an 
interaction between DE and DM -- then in general the universe evolution is no adiabatic. Thus, our considerations on thermodynamics implies that a pure $\Lambda$ component is hardly consistent with physics, unless we accept that both components -- DM and DE -- interact with each other.

\section{Effective temperature method (ETM)}

In this section we study the thermal evolution in an interaction DE/DM model using an 
effective equation of state. We follow previous works \citep{samuel} on the subject. 
Let us write the system (\ref{(21)}, \ref{(22)}) as

\begin{eqnarray}
\dot{\rho}_{x}+3H\left( 1+\omega _{x}^{eff}\right) \rho _{x} &=&0, \label{roxeq} \\
\dot{\rho}_{m}+3H\left( 1+\omega _{m}^{eff}\right) \rho _{m} &=&0, \label{romeq}
\end{eqnarray}%
where 
\begin{equation}
\omega _{x}^{eff}=\omega _{x}+\frac{Q}{3H\rho _{x}} \hspace{0.5cm}\text{and }\hspace{0.5cm}
\omega _{m}^{eff}=\omega _{m}-\frac{Q}{3H\rho _{m}},  
\end{equation}%
then according to (\ref{tda}) we can write
\begin{equation}
T\left( z\right) =T\left( 0\right) \exp \left( 3\int_{0}^{z}\frac{dz}{1+z}%
\omega ^{eff}\left( z\right) \right).  \label{(25)}
\end{equation}%
where we have used that $1+z = a_0/a$, and so for each dark component from (\ref{rox}) and (\ref{rom}) we can write
\begin{eqnarray}
T_{x}\left( z\right) &=&T_{x}\left( 0\right) \left( 1+z\right) ^{3\omega
_{x}}\exp \left[ \int_{0}^{z}d\ln \left( 1+z\right) \left( \frac{Q}{H\rho
_{x}}\right) \right] ,  \label{(28)} \\
T_{m}\left( z\right) &=&T_{m}\left( 0\right) \left( 1+z\right)
^{3\omega _{m}}\exp \left[ -\int_{0}^{z}d\ln \left( 1+z\right)
\left( \frac{Q}{H\rho _{m}}\right) \right] ,  \label{(29)}
\end{eqnarray}
and from this, we see that if interaction exists for $\omega _{m}=0$ (CDM) we get $T_{m}\left(z\right) \neq const.$ Let us apply this result to the most well known Ansatzes:  $Q=3\gamma H \rho _{x}$ and $Q=3\gamma H\rho _{m}$.

\subsection{Ansatz $Q=3\gamma H \rho _{x}$}

In this case, we get
\begin{eqnarray}
T_{x}\left( z\right) &=&T_{x}\left( 0\right) \left( 1+z\right) ^{3\left(
\omega _{x}+\gamma \right) },  \label{(36)} \\
T_{m}\left( z\right) &=&T_{m}\left( 0\right) \left( 1+z\right)
^{3\omega _{m}}\exp \left[ -3\gamma \int_{0}^{z}d\ln \left(
1+z\right) \left\{ \frac{1}{r\left( z\right) }\right\} \right] .
\label{(37)}
\end{eqnarray}%
where $r\left( z\right) =\rho _{m}\left( z\right) /\rho _{x}\left( z\right) $
is the coincidence parameter. In order to get a close solution for both temperatures we need the solution for both energy densities first. In this case -- using the equation (\ref{roxeq}) -- we get the solution
\begin{equation}
\rho _{x}(z) = \rho _{x}(0) \left( 1+z\right) ^{3\left( 1+\omega _{x}+\gamma \right)}.
\end{equation}
Using this last result in the conservation equation for dark matter $\rho_m$ in Eq.(\ref{romeq}) and using that $x=a/a_{0}$ and $d/dt\rightarrow d/da$, we can write the equation as
\begin{equation}
\frac{d\rho _{m}}{dx}+\frac{3\left( 1+\omega _{m}\right) }{x}\rho _{m}
 = 3\gamma \rho _{x}\left( a_{0}\right) x^{-\left[ 3\left( 1+\omega
_{x}+\gamma \right) +1\right] },
\end{equation}
whose solution is given by
\begin{equation}
\rho _{m}\left( x\right) =\exp \left[ -\frac{3\left(
1+\omega _{m}\right) }{x}\right] \left[ C_{2}+3\gamma \rho _{x}\left(
a_{0}\right) \int dx \exp \left[ \frac{3\left( 1+\omega _{m}\right) }{x}
\right] x^{-\left[ 3\left( 1+\omega _{x}+\gamma \right) +1\right] }%
\right]
\end{equation}
The integral can be done directly, recalling that $x=a/a_{0}=\left(
1+z\right) ^{-1}$. In this way we get $\rho _{m}\left( z\right) $ and consequently $r\left( z\right) $ y $T_{m}\left( z\right) $, according to (\ref{(37)}).

\subsection{Ansatz $Q=3\gamma H \rho _{m}$}

Using this ansatz, with $\gamma > 0$ we get
\begin{eqnarray}
T_{x}\left( z\right) &=&T_{x}\left( 0\right) \left( 1+z\right) ^{3\omega
_{x}}\exp \left[ 3\gamma \int_{0}^{z}d\ln \left( 1+z\right) r\left( z\right)  \right] ,  \label{(44)} \\
T_{m}\left( z\right) &=&T_{m}\left( 0\right) \left( 1+z\right)
^{3\left( \omega _{m}-\gamma \right) },  \label{(45)}
\end{eqnarray}%
Again, from the conservation equation (\ref{romeq}) we get
\begin{equation} \label{romsol}
\rho_{m}\left( z\right) =\rho _{m}\left( 0\right)
\left( 1+z\right) ^{3\left( 1+\omega _{m}-\gamma \right)}
\end{equation}
Using this in the conservation equation for $\rho_x$ we get
\begin{equation}
\dot{\rho}_{x}+3H\left( 1+\omega _{x}\right) \rho _{x} = -3\gamma H\rho _{m}\left( a_{0}\right) \left( a_{0}/a\right) ^{3\left( 1+\omega
_{m}-\gamma \right) }.
\end{equation}
Following the previous case we get
\begin{equation}
\rho _{x}\left( x\right) =\exp \left[ -\frac{3\left(
1+\omega _{x}\right) }{x}\right] \left[ C_{1}-3\lambda _{m}\rho _{m}\left(
a_{0}\right) \int dx\exp \left[ \frac{3\left( 1+\omega _{x}\right) }{x}
\right] x^{-\left[ 3\left( 1+\omega _{m}-\lambda _{m}\right) +1\right] }
\right]
\end{equation}
and the integral can be done explicitly. In this way knowing $\rho _{x}\left( z\right) $ and then $r\left( z\right) $ and  $T_{x}\left( z\right) $ according to (\ref{(44)}).

\section{The models}

In this section we write explicitly the formulae to be used in the study of restrictions imposed by observational data. Here we study the two models of interaction discussed in the previous section, $ Q_1 = 3\gamma H \rho_{m}$, and $ Q_2 = 3\gamma H \rho_{x}$. Both have been already studied in \cite{xia}. If $ \gamma $ is zero, then there is no interaction.

For the first model (i), the Hubble function $H(z)/H_0= E(z) $ is given by
\begin{eqnarray}\label{ezq1}
E^2(z) = \Omega_m (1 + z)^3 + \Omega_r (1 + z)^4 + \\ \nonumber
+ \Omega_{x}\left( \frac{\gamma}{w + \gamma}(1+z)^3 +
\frac{w}{w+\gamma} (1+z)^{3(1+w+\gamma)} \right),
\end{eqnarray}
where $\Omega_r = 2.469 \times 10^{-5} h^{-2}(1 + 0.2271 N_{\rm eff})$ and $N_{\rm eff}
= 3.04 $, and $ \gamma $ is the parameter that makes the interaction
manifest. Here $ \Omega_m = \Omega_c + \Omega_b $, where $\Omega_c$ is the non-baryonic 
part and $\Omega_b$ is the baryonic one.

For the second model (ii), we obtain
\begin{eqnarray}\label{ezq2}
E^2(z) = \Omega_x (1+z)^{3(1+w)} + \Omega_r (1+z)^4 + \Omega_b (1+z)^3 + \\ \nonumber
+ \Omega_{c}\left( \frac{\gamma}{w + \gamma}(1+z)^{3(1+w)} +
\frac{w}{w+\gamma} (1+z)^{3(1-\gamma)} \right).
\end{eqnarray}
Here the free parameters are $h$, $\Omega_b$, $\Omega_c$, $w$ and $\gamma$.
It is clear that for $ \gamma = 0 $ both expressions - those for models (i) and (ii) - 
reduced to that of the wCDM model.

\section{The data}

In this section we describe the five sets of observational data we have used to put constraints on the models we have defined in the previous section. Preliminary results of this work appears in \cite{dani18}. We use: measurements of the Hubble function $ H(z) $, from type Ia supernova (SNIa), baryonic acoustic oscillations (BAO), gas mass fraction in clusters $f_{gas}$ and from Cosmic Microwave Background Radiation (CMBR).

The 31 data points for the Hubble function $ H(z) $ are taken from several works and were compiled by \cite{Magana:2017nfs} expanding a redshift range from $ z = 0.07 $ to $ z = 1.965 $. In summary it comprise data points from \cite{zhang2014}, \cite{Stern2010}, \cite{moresco2012} and also from \cite{moresco2015}. It is important to notice that we have used only those $H(z)$ measurements obtained using the differential age method \cite{dem}, and we have explicitly exclude those obtained using the clustering method, because we are also using data from BAO.

The latest sample of SNIa is the Pantheon sample \cite{Scolnic:2017caz}. Here we use the chi square function defined as
\begin{equation}\label{chi2jla}
\chi^2 = (\mu - \mu_{th})^{T} C^{-1}(\mu - \mu_{th}),
\end{equation}
where $ C $ corresponds to the covariance matrix delivered in \cite{Scolnic:2017caz}, $\mu_{th} = 5 \log_{10} \left( d_L(z)/10pc\right) $ is the distance modulus where $d_L(z)$ is the luminosity distance, and the modular distance is assumed to take the shape
\begin{equation}\label{mujla}
\mu = m - M + \alpha X - \gamma Y,
\end{equation}
where $ m $ is the maximum apparent magnitude in band B, $ X $ is related to the widening of the light curves, and $ Y $ corrects the color. The cosmological parameters are then constrained along with the parameters $ M $, $ X $ and $ Y $. Also from \cite{Scolnic:2017caz} a binned version of the data was published where only $ M $ is a free parameter.

The data points we use for BAO are those compiled in \cite{evslin}. This set comprise data from the 6dF survey \cite{2011MNRAS.416.3017B} at redshift $z=0.106$, distance measurements  from \cite{Ross:2014qpa} at redshift $z=0.15$, and with data from the Baryon Oscillation Spectroscopic Survey (BOSS) at redshifts $z=0.32$, $z=0.57$ and $z=2.34$. In all these cases the baryonic peak is estimated performing an average in the radial and transverse direction. At higher redshift it is possible to measure the BAO scale in the radial and tangential directions simultaneously, proving measurements of the Hubble parameter $H(z)$ and the angular diameter distance $D_A(z)$.

The BAO observations give information about the ratio 
\begin{equation}\label{dars}
\frac{D_A(z)}{r_s} = \frac{P}{(1+z)\sqrt{-\Omega_k}}\sin \left(\sqrt{-\Omega_k}\int_0^z \frac{dz}{E(z)}\right), 
\end{equation}
for the transverse direction. Here $P=c/(r_sH_0)$ and it takes the value $30.0 \pm 0.4$ for the best $\Lambda$CDM Planck fit, $r_s$ is the co-moving sound horizon that according to Planck it takes the value $r_s=1059.68$ \cite{h0planck}, and also information about the ratio
\begin{equation}\label{dhrs}
\frac{D_H(z)}{r_s} = \frac{P}{E(z)}, 
\end{equation}
for the line-of-sight direction. The parameter $P$ was used in \cite{evslin} to perform an unanchored BAO analysis, which does not use a value for $r_s$ obtained from a cosmological constant, also performed in \cite{aubourg}.

At low redshift, because it is not possible to disentangle the BAO scale in the transverse and radial direction, the surveys give the value for the ratio $D_V(z)/r_s$, where
\begin{equation}
D_V(z) = \left[z(1+z)^2D_A(z)^2D_H(z)\right]^{1/3},
\end{equation}
which is an angle-weighted average of $D_A$ and $D_H$. From \cite{evslin} the data 
considered are: at low redshift, at $z=0.106$ we have $D_V/r_s=2.98 \pm 0.13$, and 
for $z=0.15$, $D_V/r_s = 4.47 \pm 0.17$. For high redshift we consider 
$0.00874D_H/r_s + 0.146D_A/r_s = 1.201 \pm 0.021$ and $0.0388D_H/r_s-0.0330D_H/r_s = 0.781 \pm 0.053$ at $z=0.32$; $0.0158D_H/r_s+0.101D_A/r_s = 1.276 \pm 0.011$ and $0.0433D_H/r_s-0.0368D_H/r_s = 0.546 \pm 0.026$ at $z=0.57$. Following \cite{evslin}, in order to use the BAO measurements for the Lyman $\alpha$, we used the $\chi^2$ files supplied on the website \cite{baofit} directly. In what follows, we take the Planck value for $r_s$ and use $P$ as a function of $H_0$.

Data from measurements of gas mass fraction in clusters, $f_{gas}$ was also used assuming they are sources of X-ray as suggested by \cite{Sasaki:1996zz}. In particular we use the data from \cite{Allen:2007ue} which consist in 42 measurements of the X-ray gas mass fraction $f_{gas}$ in relaxed galaxy clusters in the redshift range $0.05<z<1.1$. To determine constraints on cosmological parameters we use the model function \cite{Allen:2004cd}
\begin{equation}\label{fgas}
f_{gas}^{\Lambda CDM}(z)=\frac{b \Omega_b}{(1+0.19\sqrt{h})
\Omega_M} \left[\frac{d_A^{\Lambda CDM}(z)}{d_A(z)} \right]^{3/2},
\end{equation}
where $d_A(z)$ is the angular diameter distance, $b$ is a bias factor which accounts that the baryon fraction is slightly lower than for the universe as a whole. From \cite{eke98} it is obtained $b=0.824 \pm 0.0033$. In the analysis we also use standard priors on $\Omega_b h^2 = 0.02226 \pm 0.0023$ and $h=0.678 \pm 0.009$ \cite{PDG}.

We also use CMB data in the form of the acoustic scale $l_A$, the shift parameter $R$, and the decoupling redshift $z_{*}$. The $\chi^2$ for the CMB data is constructed as
\begin{equation}\label{cmbchi}
 \chi^2_{CMB} = X^TC_{CMB}^{-1}X,
\end{equation}
where
\begin{equation}
 X =\left(
 \begin{array}{c}
 l_A - 302.40 \\
 R - 1.7246 \\
 z_* - 1090.88
\end{array}\right).
\end{equation}
The acoustic scale is defined as
\begin{equation}
l_A = \frac{\pi r(z_*)}{r_s(z_*)},
\end{equation}
and the redshift of decoupling $z_*$ is given by \citep{husugi},
\begin{equation}
z_* = 1048[1+0.00124(\Omega_b h^2)^{-0.738}]
[1+g_1(\Omega_{m}h^2)^{g_2}],
\end{equation}
\begin{eqnarray}
g_1 & = & \frac{0.0783(\Omega_b h^2)^{-0.238}}{1+39.5(\Omega_b
h^2)^{0.763}}, \\
 g_2 & = & \frac{0.560}{1+21.1(\Omega_b h^2)^{1.81}},
\end{eqnarray}
The shift parameter $R$ is defined as in \citep{BET97}
\begin{equation}
R = \frac{\sqrt{\Omega_{m}}}{c(1+z_*)} D_L(z).
\end{equation}
$C_{CMB}^{-1}$ in Eq. (\ref{cmbchi}) is the inverse covariance
matrix,
\begin{equation}
C_{CMB}^{-1} = \left(
\begin{array}{ccc}
3.182 & 18.253 & -1.429\\
18.253 & 11887.879 & -193.808\\
-1.429 & -193.808 & 4.556
\end{array}\right).
\end{equation}
More details of the work with the data see \cite{Cardenas:2014}.

\section{Results}

For the analysis we have used the code EMCEE \cite{emcee}. It is a Python module that implement an Affine-invariant Markov chain Monte Carlo (MCMC) method. We have perform the analysis using the five data sets mentioned in the previous section. In practice we have considered a burn-in phase where we monitoring the auto-correlation time ($\tau$) and set a target number of independent samples. Then, we set 10000 MCMC steps (N) with a number of walkers in the range between 50 and 100. Our estimations of the auto-correlation times for each parameter in the three models all satisfies the relation $N/\tau \gg 50$ suggested in \cite{emcee}, a condition that is considered a good measure of assets convergence in our samplings.

The results are shown in Table \ref{tab:table01}, and the best fit plots are shown in Fig.(\ref{fig: fig1}) for model (i) and in Fig.(\ref{fig: fig2}) for model (ii). 
\begin{table}[h!]
\begin{center}
\begin{tabular}{ccccc}
\hline
\\
 \hspace{1cm} & $Q= 3H\gamma \rho_m$ & \hspace{1cm}  & $Q= 3H\gamma \rho_x$  \\
\hline
$h$            &  $0.669 \pm 0.008 $   &    &  $0.671 \pm 0.009$    \\
$\Omega_c$   &   $0.301 \pm 0.004$  &      &  $ 0.300 \pm 0.004 $  \\
$\Omega_b$   &   $ 0.049 \pm 0.001 $ &       &  $ 0.048 \pm 0.001$   \\
$\omega$   &  $-1.03 \pm 0.02$   &      &  $-1.05 \pm 0.02$    \\
$\gamma$   &   $0.071 \pm 0.006$   &   &  $ 0.07 \pm 0.005$ \\

\hline
\end{tabular}
\end{center}
\caption{Best fit values of the cosmological parameters for the interaction models using SNIa+$H(z)$+BAO+$f_{gas}+$ CMB. \label{tab:table01}}
\end{table}

As can be seen, the best fit with the full set of observational data in both models indicates positive evidence for interaction -- with a similar value for $\gamma \simeq 0.07$ -- and that the transfer of energy flows from DE to DM. Based on the considerations we have made in previous sections, we find that our theoretical constraints are in agreement with the observational evidence. We also notice that the EoS parameter for DE $\omega$ is less than $-1$ in both cases, pointing towards evidence for phantom dark energy. The rest of the parameters take best fit values that are not too different from the usual ones. The results for the best fit for model (i) is shown in Fig.(\ref{fig: fig1}). 
\begin{figure}
\includegraphics[width=15cm]{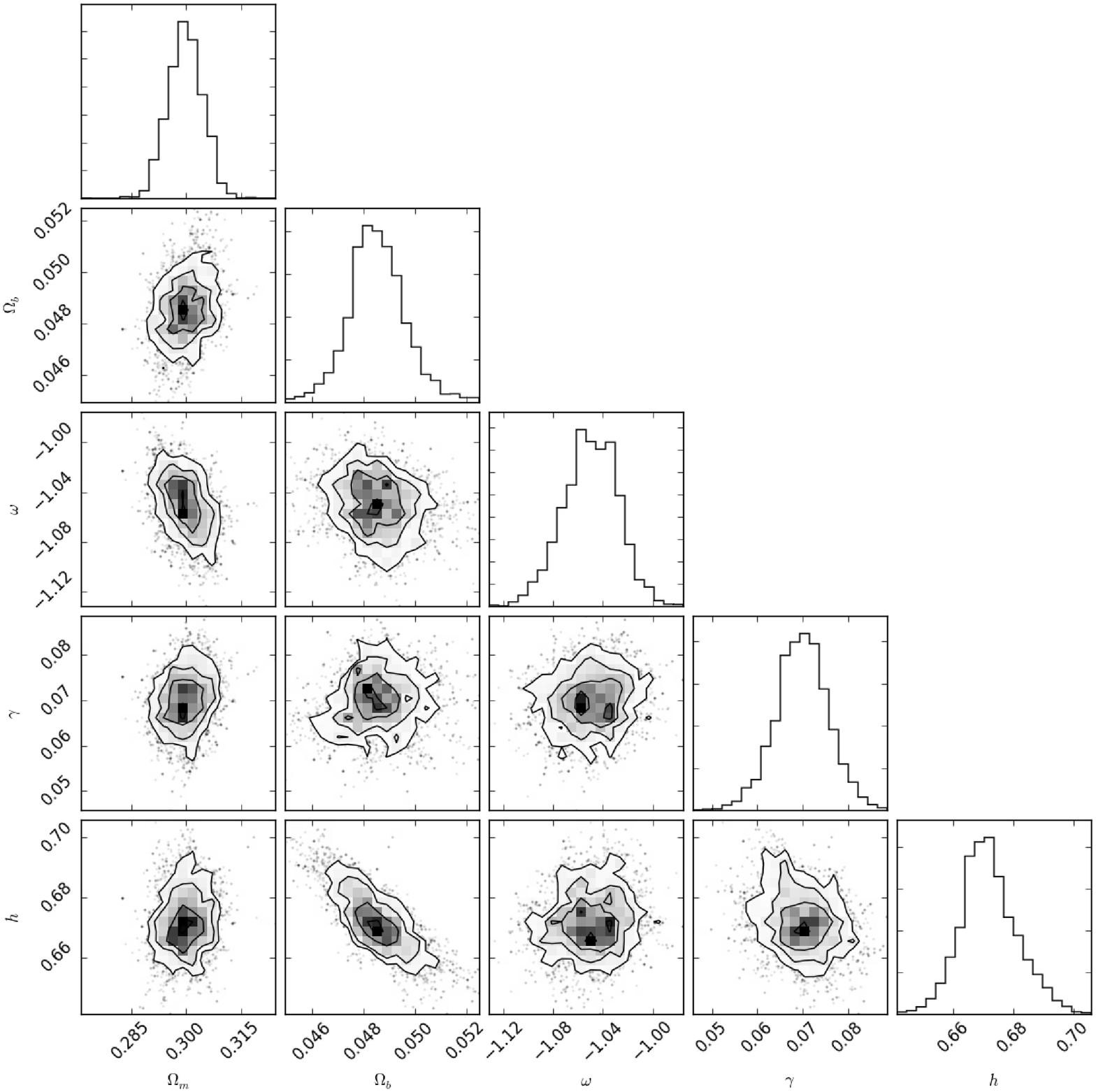}
\caption{We display the results for $ 1 \sigma $, $ 2
\sigma $ and $3 \sigma$ for the model (i) in the parameter space $( \Omega_m, \Omega_b, \omega, \gamma, h )$ using all the data.} 
\end{figure}\label{fig: fig1}

\begin{figure}[h]
\centering
\includegraphics[width=10cm]{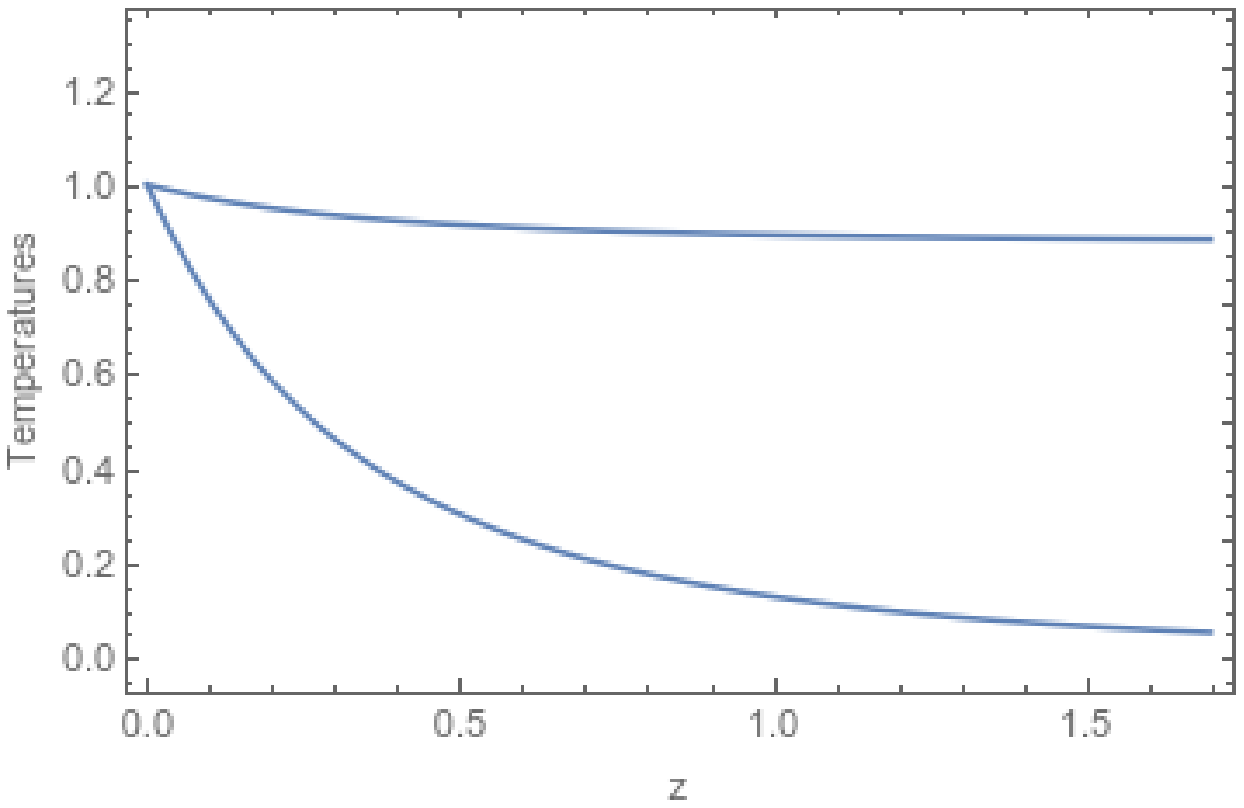}
\caption{We display the reconstructed temperatures for model (i) using the expressions (\ref{(36)}, \ref{(37)})  using all the data. The upper line is $T_m(z)/T_m(0)$ and the lower one is $T_x(z)/T_x(0)$.} \label{fig: fig3}
\end{figure}

Using these best fit values, we can use Eqs.(\ref{(36)}, \ref{(37)}) to reconstruct the temperatures for DE and DM. In Fig.(\ref{fig: fig3}) we show the result for model (i).
According to what we expect, the temperature of the dark energy grows with the expansion, and through the interaction, causes also that the temperature of dark matter to grow as well, much more moderately, but it increases. This is expected because, without interaction, DM temperature is constant, so once there is a transfer of energy from DE to DM, it will result in an increase in temperature, since the DE temperature increases with the expansion.

\begin{figure}[h]
\centering
\includegraphics[width=10cm]{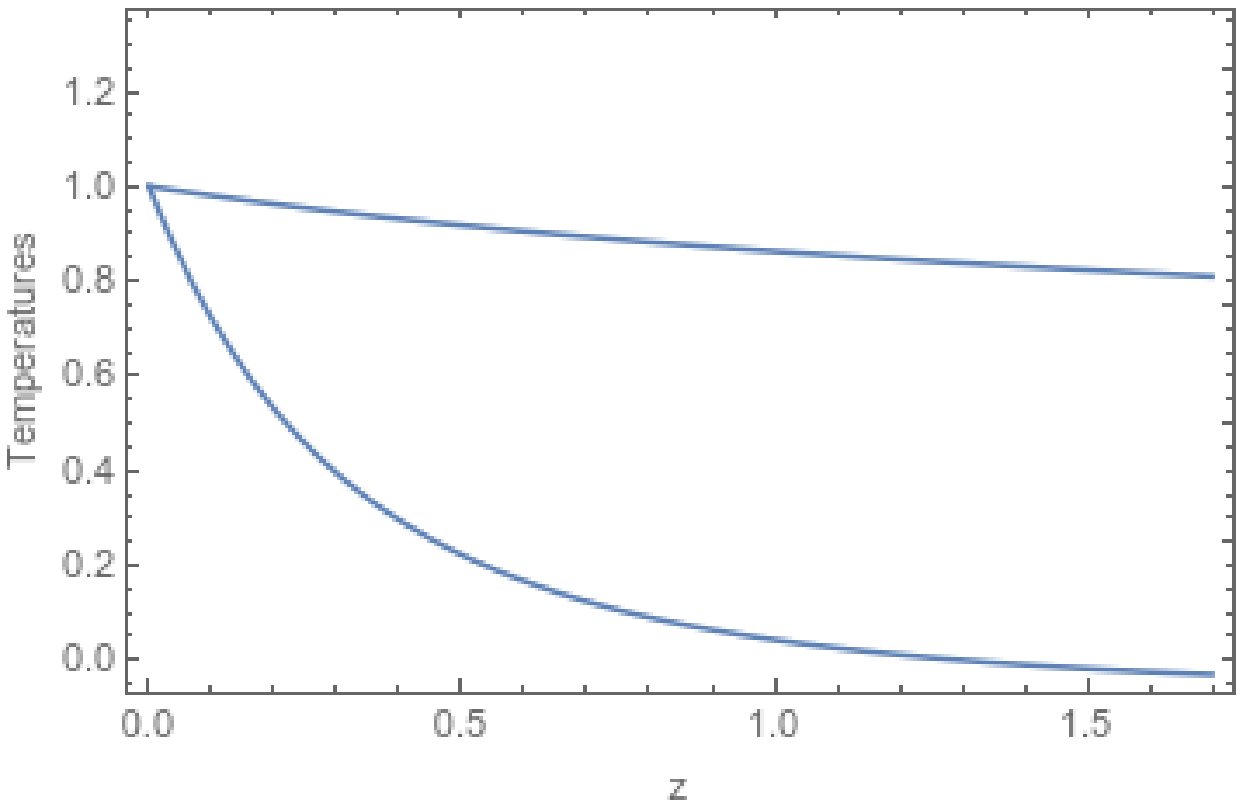}
\caption{We display the reconstructed temperatures for model (ii) using the expressions (\ref{(44)}) (\ref{(45)})  using all the data. The upper line is $T_m(z)/T_m(0)$ and the lower one is $T_x(z)/T_x(0)$.} \label{fig: fig4}
\end{figure}

\begin{figure}[h]
\centering
\includegraphics[width=15cm]{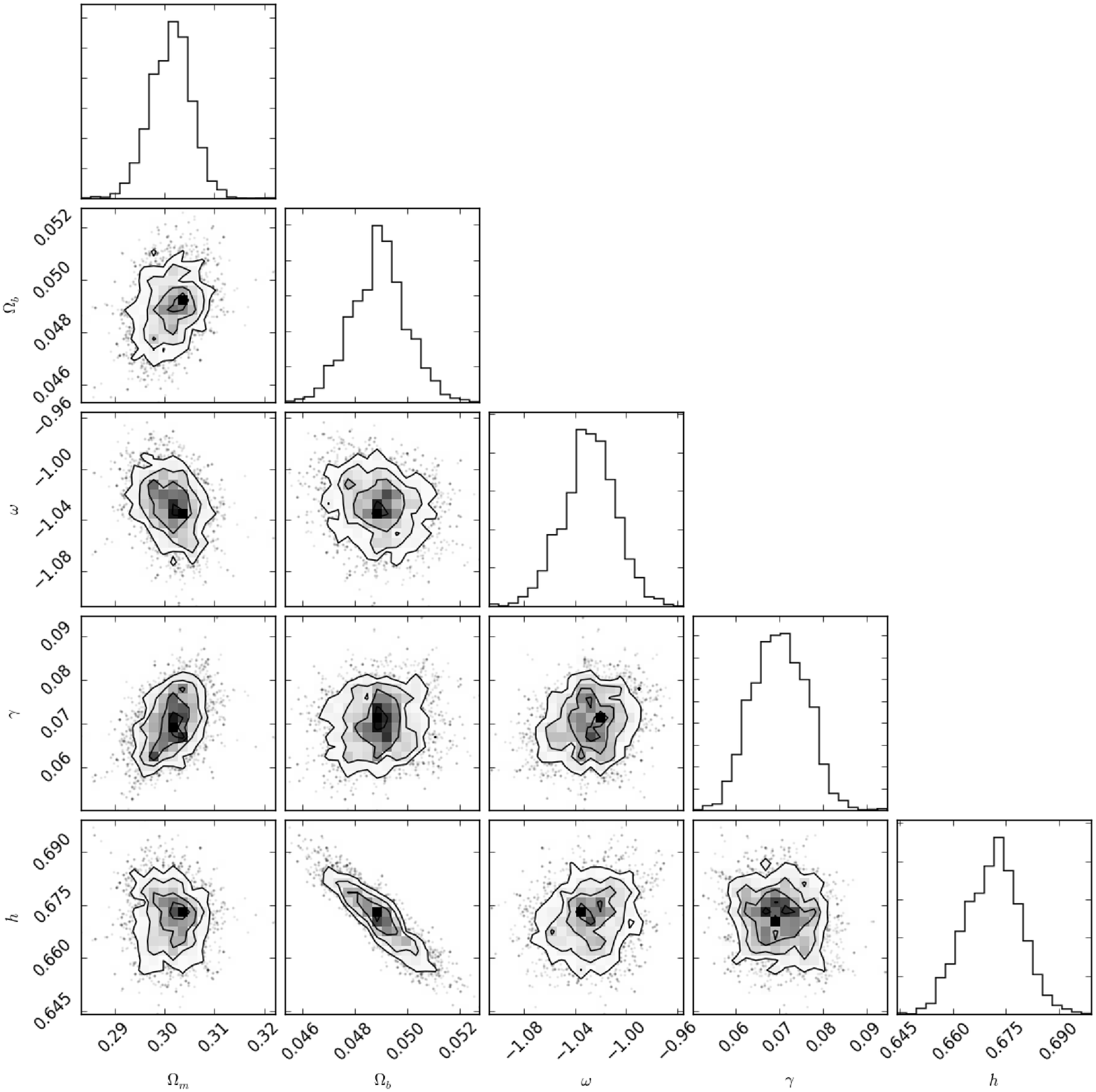}
\caption{We display confidence boundaries for $ 1 \sigma $, $ 2
\sigma $ and $ 3 \sigma $ for the model (ii) for the free parameters $( \Omega_m, \Omega_b, \omega, \gamma, h )$ using all the data.} \label{fig: fig2}
\end{figure}

In the case of model (ii) the results are very similar as can be seen in Table I. The best fit values of the parameters are shown in Fig.(\ref{fig: fig2}). The reconstructed temperatures are also very similar as can be seen in Fig.(\ref{fig: fig4}).

In this section we have performed an analysis using five geometric probes to constrain two interacting models. Compared to previous analysis as in \cite{xia}, we have used more data probes but we have obtained higher values for $\gamma$ in both cases. The reason behind this finding could be the use of BAO data at larger redshift as those at $z=2.34$ and $z=2.36$ not used in \cite{xia}. We know that the inclusion of that data points suggest strongly a departure from $\Lambda$CDM and a preference for an interaction model. It could be also the use of $f_{gas}$ data, a set with well known tension with $\Lambda$CDM \cite{Cardenas:2013roa}.


Although we have performed an analysis in the context of an interacting model, it is interesting to come back to the thermodynamic features of the LCDM model. As we have discussed here, a pure $\Lambda$ component -- understood as a source in the right hand side of Einstein's equations -- does not have any sense thermodynamically. In fact, as we have found in section II, the entropy associated to $\Lambda$ should be zero during all the universe evolution. This implies that a more sound model is necessary to explain what we are observing. Assuming that the EoS parameter evolve with redshift is equivalent to consider an interaction between components, in our case DE and DM, so, a more physical model -- thinking in thermodynamics -- must consider an interaction, something that the observational data seems to support. 


Finally, it is very difficult to measure the temperature for dark matter (see for example \cite{OPBB}, \cite{NMira}). What we know from the thermodynamic considerations is that $T_m < T_x$. Certainly, more work is needed to understand the implications of the thermodynamic evolution of our universe.


\end{document}